\documentclass[a4paper,twocolumn,aps,pra]{revtex4-1}

\newcommand{\eq}[1]{\begin{equation}#1\end{equation}}
\newcommand{\dd}{\mathrm{d}}
\newcommand{\ee}{\mathrm{e}}

\newcommand{\Tr}{\mathrm{Tr \,}}
\newcommand{\Li}[1]{\mathrm{Li}_2\left( #1 \right)}
\usepackage{amsmath}
\usepackage{amsfonts}
\usepackage{amssymb}
\usepackage{amscd}
\usepackage{graphics}
\usepackage{graphicx}
\usepackage{psfrag}
\usepackage{color}
\usepackage{colordvi}

\begin{document}

%\title{Mutual information in a nonequilibrium steady state: violation of the area law}
\title{Area law violation for the mutual information in a nonequilibrium steady state}

\author{Viktor Eisler}
\affiliation{Institute for Theoretical Physics,
E\"otv\"os Lor\'and University, P\'azm\'any s\'et\'any 1/a, 1117 Budapest, Hungary}
\author{Zolt\'an Zimbor\'as}
\affiliation{Department of Theoretical Physics, University of the Basque Country UPV/EHU, P.O. Box 644, E-48080 Bilbao, Spain}

\begin{abstract}
We study the nonequilibrium steady state of an infinite chain of free fermions,
resulting from an initial state where the two sides of the system are prepared at different temperatures.
The mutual information is calculated between two adjacent segments of the chain and is found
to scale logarithmically in the subsystem size. This provides the first example 
of the violation of the area law in a quantum many-body system outside a zero temperature regime.
The prefactor of the logarithm is obtained analytically
and, furthermore, the same prefactor is shown to govern the logarithmic increase of
mutual information in time, before the system relaxes locally to the steady state.
%from the asymptotic form of the fermionic correlation functions, invoking a Fisher-Hartwig type calculation.
%Furthermore, it is shown that the time evolution of the mutual information is also logarithmic, governed by the
%steady state prefactor.

% the same prefactor is found to govern the logarithmic increase of time-dendent mutual information
%The asymptotic state of the system has the form of a generalized Gibbs ensemble where

\end{abstract}

\maketitle

\section{Introduction}

In recent years, studies on correlations between 
subsystems in many-body states have attracted 
great attention. At the heart of these investigations is the realization
that for naturally occurring states,
the correlations  are most often restricted by an \emph{area law} \cite{ECP09}.
Historically this topic arose from black-hole physics, where the entropy of a
black hole, scaling with the area of the event horizon, was interpreted
to emerge from a general holographic principle
\cite{Bombelli86,Srednicki93, NRT09}. Later it turned 
out that similar bounds on quantum correlations, measured by the entanglement entropy,
also hold for ground states of local quantum many-body systems \cite{PEDC04, H07, BH13}.
This insight helped, among other things, to understand the power of numerical methods
capturing the structure of ground-state correlations \cite{Peschel99,Schollwoeck05}
and also led to the development of new types of trial states \cite{VCM08, CV09}.
The only relevant exceptions from a strict area law are quantum critical systems at zero temperature,
where logarithmic violations may be found \cite{Vidal03,LR09}. These are particularly well understood for
one-dimensional quantum systems with the help of conformal field theory (CFT) \cite{CC09},
but they also persist in higher dimensions for free-fermion ground states \cite{Wolf06,GK06, FZ07}.

At finite temperatures the situation is more involved, since for mixed states no unique measure
of quantum correlations exists.
Nevertheless, one can quantify the amount of \emph{total} (quantum and classical) 
correlations between two disjoint subsystems by the \emph{mutual information}. Remarkably,
this particular measure of correlations fulfills an area law for nonzero temperatures
in great generality. Namely, for any Gibbs state of a lattice system defined by a short-range Hamiltonian,
the mutual information between neighboring
subsets is proportional to the area of the common boundary \cite{WVHC08}. 
For free-fermion systems, the factor of proportionality can even be bounded by the
logarithm of the inverse temperature \cite{BKE13}. The mutual information was also
investigated numerically for Gibbs states of more general quantum \cite{MKH10,SHKM11}
and classical lattice systems \cite{WTV11,IIKM13}, with a focus on the temperature dependence
and subleading scaling behavior. 

The question naturally emerges whether such a strict area law persists
if the system is driven out of equilibrium by preparing an initial state
where two parts of the system are thermalized at different temperatures.
Particularly interesting is the case of integrable one-dimensional quantum
systems which, due to the large number of conserved quantities, do not thermalize 
in the usual sense and the steady state is given by a generalized Gibbs ensemble (GGE)
instead \cite{RDYO07,BS08,PSSV11}. For models close to an integrable point, GGE
was found to be relevant in the description of the prethermalized state \cite{KWE11}
which was also demonstrated in recent cold-atom experiments \cite{Gring12}.
However, the implications of GGE with
respect to the area law for the mutual information has not yet been addressed.

Here we demonstrate that the GGE steady state of a one-dimensional chain of
non-interacting fermions can lead to a logarithmic violation of the area law.
Due to the slow algebraic decay of the
coefficients associated with the conserved quantities in the GGE, the steady state
becomes effectively a thermal state of a long-range Hamiltonian and thus the
arguments of Ref. \cite{WVHC08} do not apply. 
From a mathematical point of view, the logarithmic growth of mutual information
with the subsystem size can be attributed to a jump
singularity in the spectral function, i.e., in the symbol of the Toeplitz matrix 
describing the fermionic correlators.
The prefactor of the logarithm will be calculated analytically using
the Fisher-Hartwig conjecture, by a generalization of the method in Ref. \cite{JK04}.
Beside determining the steady state behavior, we also study how the  mutual information
is built up in time.
It turns out that the steady-state value is reached after a logarithmic growth in time, 
the prefactor of which is given by the same one found for the steady state.

%The mutual information measures the total amount of correlations shared between
%subsystems $A$ and $B$ and is defined as
The mutual information is defined as
\eq{
I(A:B)=S(\rho_A)+S(\rho_B)-S(\rho_{AB})\, ,
\label{eq:iab}}
where $\rho_\alpha$ with $\alpha= A,B,AB$ is the \emph{reduced} density matrix
of subsystem $\alpha$ and $S(\rho_\alpha)=-\Tr \rho_\alpha \ln \rho_\alpha$
is the corresponding von Neumann entropy. The full state is defined on an infinite chain
with site indices $m \in \mathbb{Z}$ and $\rho_\alpha$ is given by the
partial trace over sites $\mathbb{Z} \backslash \alpha$. Throughout the paper we will
consider the subsystems to be neighboring segments of length $L$ with
$A=\left[ -L+1,0\right]$ and $B=\left[ 1,L\right]$.

\section{The model}

We are interested in the nonequilibrium dynamics of a free-fermion system,
resulting from an initial state given by the density matrix
\eq{
\rho_0 = \frac{1}{Z_{\ell}} \ee^{-\beta_{\ell} \mathcal{H}_{\ell}} \otimes
\frac{1}{Z_{r}} \ee^{-\beta_r \mathcal{H}_r} \, ,
\label{eq:rho0}}
which describes two disconnected reservoirs of fermions thermalized at inverse temperatures
$\beta_{\ell}$ and $\beta_r$, with $\beta_{\ell} > \beta_r$. The chemical potentials are set
to zero, corresponding to half-filling.
%We will fix the condition $\beta_r > \beta_{\ell}$. 
The respective Hamiltonians on the left and right hand
side are given by
\eq{
\begin{split}
\mathcal{H}_{\ell} &= -\frac{1}{2}\sum_{m=-\infty}^{-1}
\left(c^{\dag}_{m}c^{\phantom\dag}_{m+1} + c^\dag_{m+1}c^{\phantom\dag}_{m} \right), \\
\mathcal{H}_r &= -\frac{1}{2}\sum_{m=1}^{\infty}
\left(c^{\dag}_{m}c^{\phantom\dag}_{m+1} +c^\dag_{m+1}c^{\phantom\dag}_{m}\right),
\end{split}
\label{eq:hlr}}
where $c^{\dag}_m$ is a fermionic creation operator at site $m$.
At time $t=0$ the two semi-infinite chains are connected and the unitary time
evolution $\rho_t = \ee^{-i \mathcal{H}t} \rho_0 \ee^{i \mathcal{H}t} $
of the full system is governed by the Hamiltonian
\eq{
\mathcal{H}=-\frac{1}{2}\sum_{m=-\infty}^{\infty}
\left(c^{\dag}_{m}c^{\phantom\dag}_{m+1}  + c^\dag_{m+1}c^{\phantom\dag}_m\right) .
\label{eq:h}}

In particular, one is interested in the asymptotic behavior of the system.
The steady state $\rho_\infty$ exists if, for \emph{any} local observable $\mathcal{O}_S$
supported on a \emph{finite} set of sites $S$, the expectation values can be given as
\eq{
\lim_{t\to \infty} \Tr (\rho_t \mathcal{O}_S) = \Tr (\rho_\infty \mathcal{O}_S) \, .
\label{eq:ssdef}}
In fact, for the system at hand this steady state can be uniquely constructed
\cite{AH00,Ogata02,AP03} and reads
\eq{
\rho_{\infty} =\frac{1}{Z}\ee^{-\beta \mathcal{H}_{\mathrm{eff}}}, \quad
\mathcal{H}_{\mathrm{eff}} =
%\sum_{n=0}^{\infty} \sum_{\sigma=+,-} \mu_n^{\sigma} Q_{n}^{\sigma} ,
\sum_{n=0}^{\infty} \left( \mu_n^{+} Q_{n}^{+} + \mu_n^{-} Q_{n}^{-}\right),
\label{eq:heff}}
where $\beta = (\beta_{\ell} + \beta_r)/2$
%is the average of the initial inverse temperatures
and the effective Hamiltonian $\mathcal{H}_{\mathrm{eff}}$ involves two infinite sets
of conserved quantities \cite{GM95}
\eq{
\begin{split}
Q_{n}^{+} &= -\frac{1}{2}\sum_{m=-\infty}^{\infty}
\left( c^{\dag}_{m}c^{\phantom\dag}_{m+n} + c^{\dag}_{m+n}c^{\phantom\dag}_{m} \right) , \\
Q_{n}^{-} &= -\frac{i}{2}\sum_{m=-\infty}^{\infty}
\left( c^{\dag}_{m}c^{\phantom\dag}_{m+n} -  c^{\dag}_{m+n}c^{\phantom\dag}_{m} \right) .
\end{split}
\label{eq:qnpm}}
In particular, one has $Q_{1}^{+} = \mathcal{H}$ and $Q_{1}^{-}$ and $Q_{2}^{-}$ are,
up to a factor, the operators of the particle and energy current, respectively.
Thus, the steady state in (\ref{eq:heff}) has
exactly the form of a GGE \cite{RDYO07} with the Lagrange multipliers associated to
the conserved charges given by \cite{Ogata02}
\eq{
\mu_{n}^{+} = \delta_{n,1} \, , \quad
\mu_{n}^{-} = 
\begin{cases}
\frac{4}{\pi}\frac{\beta_{\ell}-\beta_r}{\beta_{\ell}+\beta_r}\frac{n}{n^2-1} & \text{$n$ even,}\\
0 & \text{$n$ odd.}
\end{cases}
\label{eq:munm}}
Note, that the $\mu_n^{-}$ coefficients decay asymptotically as $1/n$ and, since $Q_n^{-}$
contains hopping terms over $n$ sites, the resulting $\mathcal{H}_{\mathrm{eff}}$
is long range.

The consequences of the long-range GGE form of the steady state can also be
traced on the form of the fermionic correlation functions
\eq{
C_{mn} = %\langle c_m^{\dag} c_n\rangle =
\Tr (\rho_\infty c_m^{\dag} c_n^{\phantom\dag}) =
\int_{-\pi}^{\pi} \frac{\dd q}{2\pi} \ee^{iq(m-n)} F(q) ,
\label{eq:cmn}}
that are given by the elements of a Toeplitz matrix with a discontinuous symbol
\cite{Ogata02}
\eq{
F(q) =
\begin{cases}
\frac{1}{\ee^{\beta_r \omega_q}+1} & q \in \left( -\pi, 0 \right) \\
\frac{1}{\ee^{\beta_{\ell} \omega_q}+1} & q \in \left( 0, \pi \right)
\end{cases}
\label{eq:fq}}
where $\omega_q = -\cos q$ is the singe-particle dispersion of free fermions. 
The symbol $F(q)$ has a simple interpretation in this particle picture. Namely,
the particles with positive momenta $q>0$, initially located on the left-hand side
and propagating to the right, are described by a Fermi distribution with $\beta_{\ell}$. Similarly,
the particles with $q<0$ are emitted from the right-hand side reservoir and are thus 
thermalized at $\beta_r$.

\section{Steady-state mutual information}

The symbol in Eq. (\ref{eq:fq}) has a jump singularity at $q=0$ between the values
%$a=(\ee^{-\beta_r}+1)^{-1}=F(0^{-})$ and $b=(\ee^{-\beta_{\ell}}+1)^{-1}=F(0^{+})$
$a=F(0^{-})$ and $b=F(0^{+})$
and there is a second jump from $1-b=F(\pi)$ to $1-a=F(-\pi)$
at the ends of the spectrum. Therefore the strict proof of the area law,
worked out in Ref. \cite{BKE13} for free-fermion states with smooth symbols, cannot be
applied to this case. On the contrary, such a Fisher-Hartwig type singularity was
shown to lead to the logarithmic scaling of the entropy in the zero temperature case,
where the symbol jumps from $0$ to $1$ \cite{JK04}.

\begin{figure}[t]
\center
\includegraphics[width=\columnwidth]{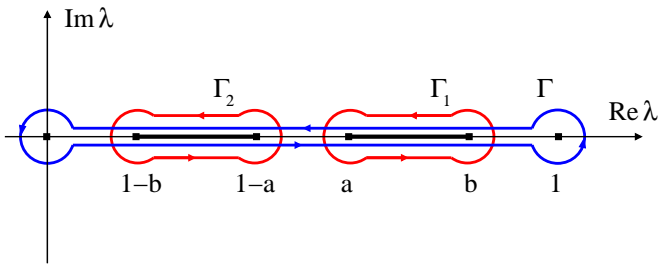}
\caption{(color online) Contour of the integral yielding the entropies $S_{\alpha}$
in the complex $\lambda$ plane. The large (blue) contour $\Gamma$ is used in
Eq. (\ref{eq:sint}), while the two small (red) contours $\Gamma_1$ and $\Gamma_2$
are associated to the jump-symbols, responsible for the $\ln L$ contributions.}
\label{fig:cont}
\end{figure}

The calculation can be generalized to obtain the steady-state mutual information $I(A:B)$.
%corresponding to the symbol $F(q)$ in Eq. (\ref{eq:fq}).
%Since one is dealing with a free fermion state,
The entropies $S_\alpha \equiv S(\rho_\alpha)$ of subsystems
$\alpha=A,B,AB$ can be written as $S_{\alpha}=\sum_k s(\lambda_{\alpha,k})$, where
\eq{
s(\lambda) = -\lambda \ln \lambda - (1-\lambda) \ln (1-\lambda) \, ,
\label{eq:sbin}}
and $\lambda_{\alpha,k}$ are the eigenvalues of the reduced correlation matrix
$\mathbf{C}_{\alpha}$ with the elements in Eq. (\ref{eq:cmn}) restricted to $m,n \in \alpha$
\cite{Vidal03}. This formula makes it possible to evaluate $I(A:B)$ numerically for
large system sizes. The analytic treatment, however, requires an
integral representation of the entropy \cite{JK04} 
\eq{
S_{\alpha}=
\frac{1}{2\pi i} \oint_\Gamma \dd \lambda \, s(\lambda) \,
\frac{\dd \ln D_\alpha(\lambda)}{\dd \lambda} \, ,\label{eq:sint}}
where $D_\alpha(\lambda) = \det (\lambda\mathbf {1}-\mathbf{C}_{\alpha})$
is a Toeplitz determinant constructed from the reduced correlation matrix $\mathbf{C}_{\alpha}$.
The contour of the integration $\Gamma$, depicted on Fig. \ref{fig:cont}, encircles the eigenvalues
$\lambda_{\alpha,k}$ of $\mathbf{C}_{\alpha}$ on the real line and,
through the logarithmic derivative of $D_\alpha(\lambda)$,
gives a pole contribution at each eigenvalue.

%%%%%%%%%%%%%%%%%%%%%%%%%%%%%%%%%%%%%%%%%%%%%%%%%%%%%%%%%%%
%
%
%%%%%%%%%%%%%%%%%%%%%%%%%%%%%%%%%%%%%%%%%%%%%%%%%%%%%%%%%%%

In order to obtain asymptotic expressions for $S_{\alpha}$, one has to
invoke the Fisher-Hartwig conjecture \cite{BT91} for the determinants $D_\alpha(\lambda)$
of Toeplitz matrices with symbol $\phi(q) = \lambda - F(q)$.
%(note that we drop the explicit $\lambda$ dependence).
First, the symbol is written in the factorized form
$\phi(q) = \psi(q) t_{\beta_1,0}(q) t_{\beta_2,\pi}(q)$ where $t_{\beta_1,0}(q)$
and $t_{\beta_2,\pi}(q)$ describe the jumps at $q=0$ and $q=\pi$, respectively,
while $\psi(q)$ is a smooth function of $q$. The canonical expressions of the
jump-singularities involve the auxiliary functions $\beta_1$ and $\beta_2$
of the variable $\lambda$ (see Appendix for definitions)
with cuts over the intervals $\left[a,b\right]$ and $\left[1-b,1-a\right]$, respectively.
The asymptotics of the $L \times L$ Toeplitz determinant is then given by
\eq{
D_{L}= \left({\cal F}[\psi]\right)^L
L^{-(\beta_1^2+\beta_2^2)}
{\cal E}[\psi,\beta_1,\beta_2] \, ,
\label{eq:fh}}
where $\mathcal{F}$, which yields the extensive part of the entropy,
is a functional of $\psi$ given by the Szeg\H{o} limit theorem \cite{BS90},
while $\mathcal{E}$ is a functional of $\psi,\beta_1,\beta_2$
and independent of $L$.
% Note, that the $\lambda$ dependence has been suppressed

The asymptotics of $S_{\alpha}$ with $\alpha=A$ or $B$
is thus evaluated through the expression in Eq. (\ref{eq:fh}), while the entropy
for the joint subsystem $\alpha=AB$ involves the determinant $D_{2L}$.
It is then straightforward to see that the extensive parts cancel out in $I(A:B)$
and the leading behavior is given by
\eq{
I(A:B) = \sigma \ln L + \mathrm{const.}
\label{eq:iab2}}
The prefactor $\sigma$ is entirely determined by the singular parts
of the symbol, described by the functions $\beta_1$ and $\beta_2$,
and thus the contour of the integration can be reduced to the loops
$\Gamma_1$ and $\Gamma_2$ encircling the cuts, see Fig. \ref{fig:cont}.
The calculation is rather lengthy and is presented in the Appendix.
The result for the prefactor reads
\begin{align}
\sigma = \frac{1}{\pi^2} \left[
a \, \Li{\frac{a-b}{a}} \right. &+ (1-a) \Li{\frac{b-a}{1-a}} \nonumber \\
+ b \, \Li{\frac{b-a}{b}} &+ \left. (1-b) \Li{\frac{a-b}{1-b}} \right] ,
%\right]
\label{eq:pref}
\end{align}
where $\Li{x}$ denotes the dilogarithm function.

\begin{figure}[t]
\center
\psfrag{L}[][][.6]{$\ell$}
\includegraphics[width=\columnwidth]{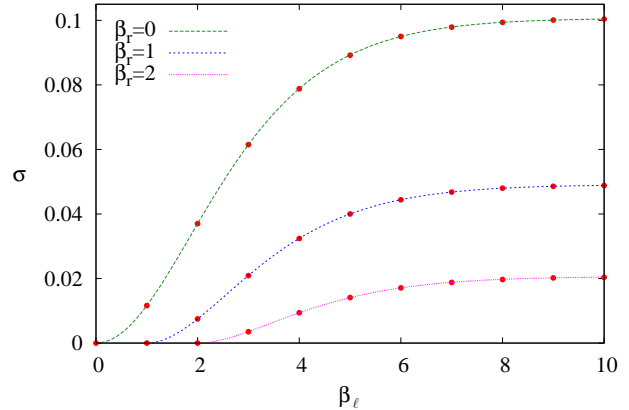}
\caption{(color online) Prefactor $\sigma$ of the logarithmic term in the steady-state mutual information
$I(A:B)$ as a function of $\beta_{\ell}$ and for various $\beta_r$. The lines represent the analytic
result in Eq. (\ref{eq:pref}), while the dots are the results of fitting the numerical data.}
\label{fig:pref}
\end{figure}

Note that the result does not depend on the details of the symbol $F(q)$ except for the values
$a=(e^{-\beta_r}+1)^{-1}$ and $b=(e^{-\beta_{\ell}}+1)^{-1}$
at the $q=0$ discontinuity,
and is manifestly symmetric under the simultaneous exchange $a \to 1-a$ and $b \to 1-b$.
Comparing with the numerical results, obtained from logarithmic fits on data sets up to
segment sizes $L=200$, one finds an excellent agreement with a precision up to four digits,
as shown in Fig. \ref{fig:pref}.

%%%%%%%%%%%%%%%%%%%%%%%%%%%%%%%%%%%%%%%%%%%%%%%%%%%%%%%%%%%
%
%
%%%%%%%%%%%%%%%%%%%%%%%%%%%%%%%%%%%%%%%%%%%%%%%%%%%%%%%%%%%

\section{Dynamics of mutual information}

The next question we address is how the steady-state value of the mutual
information is reached in the course of the time evolution, after the two
sides of the system are connected. This can be considered as a 
generalization of the local quench setup at zero temperature where
the time evolution of entanglement entropy was studied \cite{EP07,CC07}.
Since the initial state in Eq. (\ref{eq:rho0}) is factorized, one clearly has $I_0(A:B)=0$.
To study the growth of the mutual information, one needs the time dependent fermionic
correlations \cite{EP07}
\eq{C_{mn}(t)=i^{n-m} \sum_{k,l \in \mathbb{Z}} i^{k-l} J_{m-k}(t) J_{n-l}(t) C_{kl}(0) \, ,
\label{eq:cmnt}}
where $J_m(t)$ are Bessel functions and the initial correlation matrix is given by
$C_{kl}(0) = \Tr (\rho_0 c^{\dag}_k c_l^{\phantom\dag})$. Due to exponentially vanishing contributions
from terms with $|m-k| \gg t$ and $|n-l| \gg t$, the infinite sums in Eq. (\ref{eq:cmnt})
can be truncated and the matrix elements $C_{mn}(t)$ can be evaluated numerically.
The mutual information $I_t(A:B)$ can then be extracted from a formula analogous to Eq.
(\ref{eq:iab}) by diagonalizing the reduced correlation matrices $\mathbf{C}_\alpha(t)$ and
using $S_{\alpha}(t)=\sum_k s(\lambda_{\alpha,k}(t))$.

\begin{figure}[t]
\center
\includegraphics[width=\columnwidth]{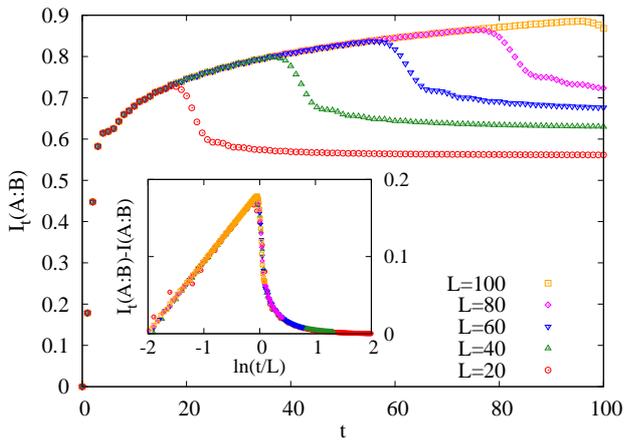}
\caption{(color online) Time evolution of the mutual information $I_t(A:B)$ for
$\beta_{\ell}=5$, $\beta_r=0$ and various segment sizes $L$, after connecting the chains at $t=0$.
The inset shows the difference from the steady state value $I (A:B)$, plotted against $\ln(t/L)$.}
\label{fig:iabt}
\end{figure}

The resulting $I_t(A:B)$ is shown on Fig. \ref{fig:iabt} for a range of segment sizes $L$
with $\beta_{\ell}=5$ and $\beta_r=0$ fixed. After an initial logarithmic increase, the mutual
information drops sharply around $t=L$ and converges slowly towards its steady-state
value $I(A:B)$. Considering the distance from this asymptotic value, the curves
for different $L$ can be scaled together using the variable $\ln (t/L)$, which is shown
in the inset of Fig. \ref{fig:iabt}. One can see a cusp emerging between the growth
and relaxation parts of the scaling function, with the former showing a pure logarithmic
behavior. The prefactor of the logarithm was fitted for various values of $\beta_{\ell}$
and $\beta_r$ and, with a good precision, we recover the steady-state prefactor in Eq.
(\ref{eq:pref}).

%%%%%%%%%%%%%%%%%%%%%%%%%%%%%%%%%%%%%%%%%%%%%%%%%%%%%%%%%%%
%
%
%%%%%%%%%%%%%%%%%%%%%%%%%%%%%%%%%%%%%%%%%%%%%%%%%%%%%%%%%%%

The appearance of the same prefactor governing the time evolution as well as the
steady state behavior is reminiscent of the situation for the entanglement entropy
in a local quench at zero temperature. In the latter case, for $t \ll L$, one has
$S \sim 1/3 \ln(t)$ and thus the equilibrium scaling appears with $t$ and $L$ interchanged
\cite{EP07,CC07}. However, for intermediate times one has additional terms in the entropy,
obtained from a CFT calculation and scaling as $\ln (L \pm t)$ \cite{CC07},
which are not present for $I_t(A:B)$.
We have also checked the dynamics of the mutual information on a finite chain
of length $2L$ where the same logarithmic growth of $I_t(A:B)$ persists up to $t \approx 2L$,
in contrast to the zero temperature case, where the entropy reaches a maximum at $t=L$
\cite{SD11}.

\section{Conclusions and outlook}

In conclusion, we have shown that the area law for the mutual information breaks
down in a simple nonequilibrium steady state of free fermions.
Remarkably, all the previous examples of local Hamiltonians producing such logarithmic violations
are essentially restricted to the zero temperature regime. This includes, on one hand, the ground \cite{LR09,CC09}
and low-lying excited states \cite{M09,ABS11} of critical systems, as well as highly excited \emph{pure} states of
free fermions which, however, can be interpreted to be ground states of effective local Hamiltonians \cite{AFC09}.
A simple example is given by the pure current-carrying steady state of
the XX chain \cite{EZ05}. This formally corresponds to a GGE of Eq. (\ref{eq:heff}) in the limit $\beta \to \infty$ with only
two nonzero multipliers $\mu_1^+$ and $\mu_2^{-}$, defining
the effective local Hamiltonian it is the ground state of.

In contrast, here we have pointed out the logarithmic scaling of the mutual information
in a clearly nonzero temperature context, providing the first violation of
mixed-state area laws \cite{WVHC08,BKE13,KE13}.
The necessary condition for the violation is the slow algebraic decay of the Lagrange multipliers $\mu^{-}_n$ in the GGE which,
however, does not seem to be a sufficient one.
Indeed, a similar long-range behavior has recently been pointed out for a magnetic field quench in the transverse
Ising chain, where the multipliers $\mu^{+}_n$, associated to analogous conserved charges, were shown
to decay as $1/n$ \cite{FE13}. Nevertheless, the symbol of the respective (block-Toeplitz) 
correlation matrix does not show any jump singularities in this case, and thus the resulting extensive subsystem entropies
do not involve logarithmic corrections. A similar conclusion was reached by a recent analytic calculation of
the entropy after an interaction quench in a Bose gas where the subleading term evaluates to a constant \cite{CKC13}.
Hence it is an interesting open problem whether a global quench without time-reversal symmetry breaking
could produce a GGE with a mutual information asymptotics that violates the area law.

On the other hand, by following our result, various such violations 
may be found among the nonequilibrium steady states.
In particular, the initial condition in Eq. (\ref{eq:rho0}) can be considered for
the XY model, leading to a steady state where the spin-correlation matrices have
a block-Toeplitz form with discontinuous symbols \cite{AP03}. Presumably, this would
lead again to logarithmic violations of $I(A:B)$ and the analytical calculation might
even be generalized to this case. One could also address the question, whether
long-range spin-correlations, that are also common features of nonequilibrium steady
states via incoherent driving \cite{PP08,PZ10}, could alone be responsible for an
area-law violation in general quantum chains.
Finally, it would be interesting to see if the calculations can be carried through in the framework of
nonequilibrium CFT, where the corresponding steady states have recently been constructed \cite{BD12}.
Such an approach might shed light to some universal aspects of the problem.

\begin{acknowledgments}
The work of V.E. was realized in the framework of T\'AMOP 4.2.4.A/1-11-1-2012-0001
``National Excellence Program''. The project was supported by the European Union
and co-financed by the European Social Fund. Z.Z. acknowledges funding by the Basque
Government (Project No. IT4720-10) and by the European Union through the
ERC Starting Grant GEDENTQOPT and the CHIST-ERA QUASAR project.
\end{acknowledgments}

\appendix*
\section{Analytical Calculation of $I(A:B)$}

In this Appendix, the complete analytical derivation of
the logarithmic scaling of the mutual information $I(A:B)$ is presented,
providing a closed form for the prefactor of the logarithm. In the calculation
we generalize the method of Ref. \cite{JK04}.

\begin{figure}[t]
\center
\psfrag{L}[][][.6]{$\ell$}
\psfrag{R}[][][.6]{r}
\includegraphics[width=\columnwidth]{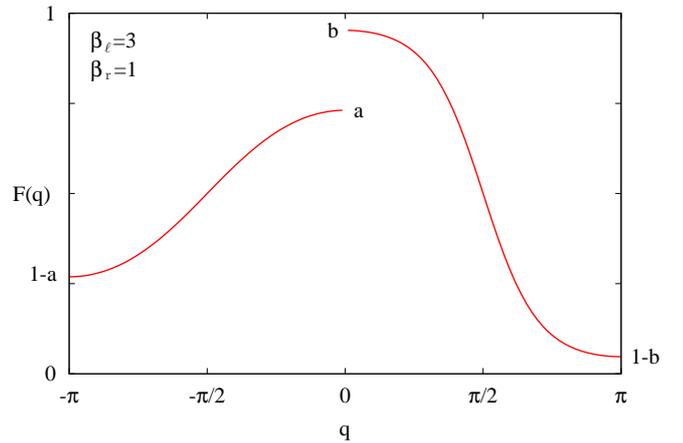}
\caption{Sketch of the symbol $F(q)$ for a given $\beta_{\ell}$ and $\beta_r$.}
\label{fig:fq}
\end{figure}

As discussed in the main text, the von Neumann entropy of $L$ consecutive spins
in the translational invariant steady state $\rho_\infty$ is given by $S_L= \sum_{k=1}^L s(\lambda_k)$,
where $s(\lambda) =- \lambda \ln \lambda - (1-\lambda) \ln (1- \lambda)$.
The $\lambda_k$ are eigenvalues of a Toeplitz matrix corresponding to
the symbol $F(q)$ defined in Eq. (\ref{eq:fq}) and sketched on Fig. \ref{fig:fq}.
The mutual information of two adjacent subsystems of length $L$ is then given
by $I(A:B)=2S_L -S_{2L}$. Using the residue theorem, we can rewrite 
the entropy as
\begin{align}
S_L = \sum_{k=1}^{L} s(\lambda_{k})&= \frac{1}{2\pi i}\oint_{\Gamma}  {\rm d} \lambda \,
s(\lambda) \sum_{k=1}^L \frac{1}{\lambda - \lambda_{k}} \nonumber \\
&=\frac{1}{2\pi i}\oint_{\Gamma} {\rm d}\lambda \,
s(\lambda)\, \frac{{\rm d}\ln D_L(\lambda)}{{\rm d}\lambda}\, ,
\label{contint}
\end{align}
where the contour $\Gamma$ is shown  in Fig. \ref{fig:cont} and  $D_L(\lambda)$ is the
determinant of the $L \times L$ Toeplitz matrix $T_L$ with entries
\begin{equation}
(T_L)_{kl}= \int_{-\pi}^{\pi} \frac{\dd q}{2\pi} \ee^{iq(k-l)} \phi(q) \, ,
\end{equation}
generated by the symbol $\phi(q)=\lambda - F(q)$.

To calculate $D_L = \det (T_L)$, we use a simplified version of the Fisher-Hartwig conjecture
\cite{Basor78, BT91}:
Suppose that $\phi(q)$ has the following factorization form
\begin{equation}
\phi(q)=\psi(q) \prod_{j=1}^R t_{\beta_j, \,q_j}(q) \, , 
\end{equation}
where $\psi(q)$ is a continuously  differentiable function and $t_{\beta_j,\,q_j}$ describe
jumps at positions $q=q_j$ in the form
\begin{equation}
t_{\beta_j,\,q_j}(q)=\exp [-i\beta_j (\pi{-}q{+}q_j)],
\end{equation}
where the $2\pi$ periodic quasi-momenta $q$ are taken from the interval 
$q_j<q <2\pi{+}q_j$.
Then the $L \to \infty$ asymptotics of the determinant is
\begin{equation}
D_L= \left({\cal F}[\psi]\right)^{L}
\left(\prod_{j=1}^R L^{-\beta_j^2}\right)
{\cal E}[\psi, \{\beta_j\},\{q_j\}]
, \label{fh}
\end{equation}
where ${\cal F}[\psi]=\exp \left(\frac{1}{2\pi} \int_0^{2\pi}\ln
\psi(q) \mathrm{d} q \right)$, and the ${\cal E}$ term does not depend on $L$.

%%%%%%%%%%%%%%%%%%%%%%%%%%%%%%%%%%%%%%%%%%%%%%%%%%%%%%%%%%%
%
%
%%%%%%%%%%%%%%%%%%%%%%%%%%%%%%%%%%%%%%%%%%%%%%%%%%%%%%%%%%%

In our case, the symbol is $\phi(q)=\lambda - F(q)$ and there are two jumps, hence 
the canonical factorization reduces to 
$\phi(q)=\psi(q) t_{\beta_1,0}(q) t_{\beta_2,\pi}(q)$, where 
\begin{align}
&\beta_1(\lambda)=\frac{1}{2\pi i} \ln \left( \frac{\lambda -(e^{-\beta_r}+1)^{-1}}{\lambda-(e^{-\beta_{\ell}}+1)^{-1}}\right) ,\\
&\beta_2(\lambda)=\frac{1}{2\pi i} \ln \left( \frac{\lambda -(e^{\beta_{\ell}}+1)^{-1}}{\lambda- (e^{\beta_r}+1)^{-1}}\right) .
\end{align}
The logarithm of $D_L$ reads
\begin{equation}
\ln D_{L}=  L \ln \mathcal{F}[\psi]   - (\beta^2_1(\lambda) + \beta^2_2(\lambda))
\ln L + \ln \mathcal{E} .
\end{equation}
Since we are interested only in the leading behavior of $I(A:B)$, we drop the last term
which gives a $\mathcal{O}(1)$ contribution. The derivative of $\ln D_{L} $ then reads
\begin{align}
\frac{\mathrm{d} \ln D_{L}(\lambda)} {\mathrm{d} \lambda} &= 
\frac{\mathrm{d} \ln \mathcal{F}[\psi]} {\mathrm{d} \lambda}  L 
-\frac{a-b}{ \pi i}\left[ \frac{\beta_1(\lambda)}{(a-\lambda)(b-\lambda)} \right.\nonumber \\
 & \left. \phantom{=} +  
\frac{\beta_2(\lambda)}{(1-a-\lambda)(1-b-\lambda)}  \right]\ln L \, , 
\end{align}
where $a=(e^{-\beta_r}+1)^{-1}$ and $b=(e^{-\beta_{\ell}}+1)^{-1}$.

According to Eq. (\ref{contint}), this has to be integrated along the large contour $\Gamma$
depicted on Fig. \ref{fig:cont} of the main text, containing the interval $[0,1]$.
Let us here emphasize that the extensive (linear in $L$) part of $S_L$
has indeed contributions from the entire contour $\Gamma$.
However, it is easy to see that in $I(A:B)$ the term proportional to $L$ drops out
and thus only a part of the contour is of importance.
We will show this using the fact that in the neighborhood of the real line one has
\begin{align}
\beta_1(x+i 0^{\pm})&=\frac{1}{2\pi i} \left[\ln \frac{a-x}{b-x}
\mp i(\pi-0^+)\right] \nonumber \\
&=\beta_1(x) \mp \left(\tfrac{1}{2} - 0^+\right),
\label{b1}
\end{align}
for $x \in (a,b)$ and similarly for $\beta_2$:
\begin{align}
\beta_2(x+i 0^{\pm})&=\frac{1}{2\pi i} \left[\ln \frac{(1-b)-x}{(1-a)-x}
\mp i(\pi-0^+)\right] \nonumber \\
&=\beta_2(x) \mp \left(\tfrac{1}{2} - 0^+\right).
\label{b2}
\end{align}
with $x \in (1{-}b,1{-}a)$. In other words the functions $\beta_1$ and $\beta_2$
have cuts along the intervals $[a,b]$ and $[1{-}b, 1{-}a]$, respectively, but they are analytic
on the rest of $[0,1]$. This means that we can reduce the contour integration along $\Gamma$
to the contours $\Gamma_1$ and $\Gamma_2$ that encircle these cuts, see Fig. \ref{fig:cont}
in the main text. Thus, we obtain that $I(A:B)=\sigma \ln L +  \mathrm{const}$, with
\begin{align}
\sigma= \frac{a{-}b}{2 \pi^2} & \left[ \oint_{\Gamma_1} \mathrm{d} \lambda
\frac{s(\lambda) \, 
\beta_1(\lambda)}{(a{-} \lambda)(b-\lambda)} + \right. \nonumber \\ 
&\left. \,\oint_{\Gamma_2} \mathrm{d} \lambda
\frac{s(\lambda) \, 
\beta_2(\lambda)}{(1-a- \lambda)(1-b-\lambda)}\right] .
\label{cint12}
\end{align}

\begin{figure}[t]
\center
\includegraphics[width=\columnwidth]{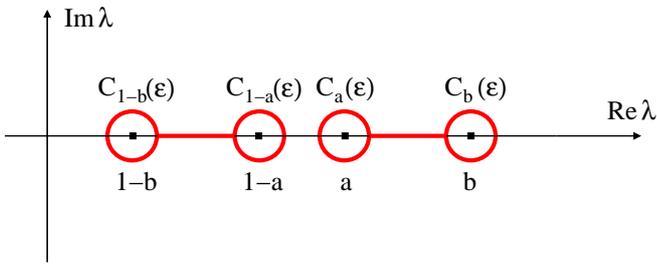}
\caption{The final integration contour decomposes into two intervals and
four circular contours around the points $a$, $b$, $1-a$ and $1-b$.}
\label{fig:Fin-Cont}
\end{figure}

The contours $\Gamma_1$ and $\Gamma_2$ can now be contracted and, using Eqs. (\ref{b1}) and (\ref{b2}),
the integration has to be carried out on the intervals $[a + \epsilon,b - \epsilon]$ and $[1-b + \epsilon, 1-a - \epsilon]$
and along circular contours around the points $a$, $b$, $1-a$ and $1-b$ , see Fig. \ref{fig:Fin-Cont}.
A further simplification occurs by observing the symmetry of the problem under the
exchange of variables $\lambda \to 1-\lambda$. Indeed, one has $\beta_2(1-\lambda)=-\beta_1(\lambda)$
where the minus sign cancels out with the reversal of the direction $\Gamma_2 \to -\Gamma_1$ of the contours
upon reflection. Thus the two contributions in Eq. (\ref{cint12}) are equal and lead to the following sums of integrals
%
%%%%%%%%%%%%%%%%%%%%%%%%%%%%%%%%%%%%%%%%%%%%%%%%%%%%%%%%%%%
%
%
%%%%%%%%%%%%%%%%%%%%%%%%%%%%%%%%%%%%%%%%%%%%%%%%%%%%%%%%%%%
%
\begin{align}
 \sigma&= \lim_{\epsilon \to 0} \frac{a{-}b}{\pi^2} 
\left[   \int_{a+\epsilon}^{b - \epsilon}  \frac{ {\rm d} \lambda \,  s(\lambda)}{(a-\lambda)(b-\lambda)}  +
\right. \nonumber\\
& \left.  \oint_{C_a(\epsilon)} 
\frac{\mathrm{d} \lambda \, s(\lambda)  \beta_1(\lambda)}{(a- \lambda)(b-\lambda)}  
 {+} \oint_{C_b(\epsilon)}  
\frac{\mathrm{d} \lambda\, s(\lambda)  \beta_1(\lambda)}{(a- \lambda)(b-\lambda)} \right] , 
\label{eq:mut_inf_final_int} 
\end{align}
where $C_{v}(\epsilon)$ denotes a circular contour of radius $\epsilon$ with $v=a,b$ as center.
Let us evaluate such a principal value integral around $b$.
Substituting $\lambda=b+\epsilon \ee^{i\theta}$ one has
\begin{align}
& \lim_{\epsilon \to 0} \oint_{C_b(\epsilon)} \frac{{\rm d} \lambda}{2\pi i} s(\lambda)
\frac{\ln(\lambda - a)-\ln(\lambda-b)}{(\lambda -a)(\lambda -b)}= \nonumber \\
& \lim_{\epsilon \to 0}  \int_{-\pi}^{\pi} \frac{{\rm d} \theta}{2\pi} s(b)
\frac{\ln(b-a) -  \ln (\epsilon) -i \theta}{b-a}= \nonumber\\
& \lim_{\epsilon \to 0}\frac{s(b)}{b-a} \ln\left(\frac{b-a}{\epsilon}\right) ,
\label{cint}
\end{align}
where in the second line we used $\dd \lambda = i\epsilon \ee^{i\theta}\dd \theta $
which cancels out the term $(\lambda-b)$ in the denominator.
Note, that the result is divergent and one has to consider it as a limit. The integral
for $C_a(\epsilon)$ is evaluated analogously with the substitution
$\lambda = a-\epsilon \ee^{i\theta}$ and yields a similar result where $s(b)$ is replaced
with $s(a)$.

For the line-integral, we get the following expression
\begin{align}
&\lim_{\epsilon \to 0} \int_{a{+}\epsilon}^{b{-} \epsilon} \frac{ {\rm d} \lambda \,
s(\lambda)}{(a{-}\lambda)(b{-}\lambda)} =\nonumber \\
 \frac{1}{a-b}  &\left[  \; a \, \Li{\frac{a{-}b}{a}} \right. + (1{-}a) \Li{\frac{b{-}a}{1{-}a}} \nonumber \\
&+ b \, \Li{\frac{b{-}a}{b}}  + \left. (1{-}b) \Li{\frac{a{-}b}{1{-}b}} \,\right] \nonumber \\
+&  \lim_{\epsilon \to 0}\frac{s(a) + s(b)}{b-a} \ln\left(\frac{\epsilon}{b-a}\right) ,
\label{lint}
\end{align}
where $\Li{x}$ is the dilogarithm function defined as
\begin{equation}
\Li{x} = -\int_{0}^{x} \dd \lambda \, \frac{\ln (1-\lambda)}{\lambda} \, .
\end{equation}
Note, that the last term of Eq. (\ref{lint}) is again divergent. However, inserting it
into Eq. (\ref{eq:mut_inf_final_int}) together with the result (\ref{cint}) for the circular contours,
the divergences cancel out and one finally obtains the prefactor $\sigma$ as given
in Eq. (\ref{eq:pref}) of the main text.

\bibliographystyle{apsrev4-1.bst}

\bibliography{mutinfness_refs}

%\clearpage
%\bibliography{mutinfness_refs_supp}

\end{document}